\documentclass[twocolumn,secnumarabic,amssymb,amsmath,nofootinbib,tightenlines,nobibnotes,aps,prl,superscriptaddress,10pt]{revtex4}

\usepackage{amsmath}%
\usepackage{longtable}%
\usepackage{bm}%
\usepackage{epsfig}

\newtheorem{theorem}{Theorem}

\newtheorem{lemma}[theorem]{Lemma}

\newcommand{\rd}{{\rm d}}
\newcommand{\be}{\begin{equation}}
\newcommand{\ee}{\end{equation}}
\newcommand{\bey}{\begin{eqnarray}}
\newcommand{\eey}{\end{eqnarray}}

\newcommand{\lan}{\langle}
\newcommand{\ra}{\rangle}

\newcommand{\ph}{\varphi}

\newcommand{\e}{\varepsilon}

\newcommand{\bR}{{\mathbb R}}

\newcommand{\bZ}{{\mathbb Z}}

\newcommand{\ann}{a^{\hphantom{+}}}
\newcommand{\cre}{a^{+}}

\newcommand{\wh}{\widehat}

\newcommand{\cF}{{\cal F}}

\newcommand{\cH}{{\cal H}}

\input epsf

\newcommand{\donothing}[1]{}

\begin{document}

\title{The ground state energy of  a low density Bose gas: a second order upper bound}
\author{L\'aszl\'o Erd\H os}
\author{Benjamin Schlein}
\affiliation{Institute of Mathematics, University of Munich,
Theresienstr. 39, D-80333 Munich, Germany}
\author{Horng-Tzer Yau}
 \affiliation{Department of Mathematics, Harvard University,
One Oxford Street, MA 02138, USA}
\date{15 August 2008}

\begin{abstract}
Consider $N$  bosons in a finite box $\Lambda= [0,L]^3\subset \bR^3$
interacting via a two-body  nonnegative soft
potential $V= \lambda \tilde V$
with $\tilde V$ fixed and $\lambda>0$ small.
We will take the limit $L, N \to \infty$ by keeping the density
  $\varrho= N/L^{3}$ fixed
and small. We construct a variational state which gives an
 upper bound on the ground state energy
per particle $\e$
$$
\e \le  4\pi\varrho a \Big [1+
\frac{128}{15\sqrt{\pi}}( \varrho a^3)^{1/2}S_\lambda \Big ]
 + O(\varrho^2|\log\varrho| ), \quad \mbox{as $\varrho\to 0$}
$$
with a constant satisfying
$$
   1\leq S_\lambda \leq 1+C\lambda \; .
$$
Here $a$ is the scattering length of $V$ and thus depends on $\lambda$.
In comparison, the prediction by Lee-Yang \cite{LYang} and  Lee-Huang-Yang \cite{LHY}
asserts that $S_\lambda=1$ independent of $\lambda$.

\bigskip


\bigskip

\end{abstract}

\maketitle

\section{Introduction}

Although Bose-Einstein condensation has been firmly established
since the experiments \cite{Kett, CW}, rigorous understanding of the Bose
gas starting from the many-body Schr\"odinger equation is still in a very
rudimentary
stage and many  theoretical predictions  at present
are still based on uncontrolled approximations. Notable exceptions are
the rigorous derivations of the Gross-Pitaevskii equation in both
 the stationary and the dynamical settings \cite{LSY, ESY}. In particular, in the
limit of low density, the leading term of the ground state energy
per particle of an interacting Bose gas was proved by Dyson (upper
bound) \cite{D} and Lieb-Yngvason (lower bound) \cite{LY} to be $4\pi a \varrho$
where $a$ is the scattering length of the two-body
potential and $\varrho$ is the density. The famous second order
correction to this leading term was first computed by Lee-Yang \cite{LYang}
 (see also Lee-Huang-Yang \cite{LHY} and  the recent paper
by Yang \cite{Y} for results in other dimensions).
In this paper,   we construct a trial function with a second
order term in the energy which,  up to a constant factor, is the same as  predicted in
 \cite{LYang, LHY}.  To present this result, we now introduce  our setup rigorously.

Consider $N$ interacting bosons in a finite box $\Lambda= [0,L]^3\subset \bR^3$
with periodic boundary conditions.  Let $ \tilde V$ be a smooth,
radially symmetric, nonnegative  potential  with fast decay.
The two-body interaction $V$ is given by $V= \lambda \tilde V$ with
$\lambda$ a small constant.
We will first take the limit $L, N \to \infty$ by keeping the density
$\varrho= N/L^{3}$ fixed. In the
limit $\varrho \to 0$, the leading term for the ground state energy per particle
of this system  is $4 \pi  \varrho a$, where $a$ is the scattering length
of the potential $V$. The Lee-Yang's prediction of the energy per particle
up to the second order term is given by
\be\label{nextorder}
\e = 4\pi\varrho a \Big [1+
\frac{128}{15\sqrt{\pi}}( \varrho a^3)^{1/2}
 + \cdots  \Big ] \; .
\ee
The approach by Lee-Yang \cite{LYang} is based on the pseudo-potential approximation \cite{HY, LHY}
and the ``binary collision expansion method'' \cite{LHY}. One can also obtain \eqref{nextorder} by performing the
Bogoliubov \cite{B} approximation and then replacing the integral of the potential
by its scattering length \cite{Landau}.  Another derivation of \eqref{nextorder}
was later given  by Lieb \cite{L} using
a self-consistent closure assumption  for the hierarchy of correlation functions.
Although  these approaches gave the same answer for
the second order term \eqref{nextorder}, it is nevertheless difficult to extract rigorous
results on the energy using these ideas.  In fact, the first rigorous
upper bound on the energy to the leading order  by  Dyson \cite{D} was
 based on a completely different construction.  The  proof of Lieb-Yngvason
 \cite{LY} on the lower bound of the energy to the leading order was also
very different from the earlier approaches.

The trial wave function of Dyson \cite{D} also shows that the next order
correction in energy for hard core bosons
is bounded from above by $ C (\varrho a^{3})^{1/3}$. The same upper bound
for soft potentials was obtained in \cite{LSY}.
In this paper, we construct a variational state which gives a rigorous
 upper bound on the ground state energy
per particle
\be
\e \le 4\pi\varrho a \Big [1+
\frac{128}{15\sqrt{\pi}}( \varrho a^3)^{1/2}S_\lambda \Big ]
 + O(\varrho^2|\log \varrho| )  \;
\label{E-1}
\ee
with $S_\lambda \leq 1+C\lambda$.
The second order term in   this upper bound \eqref{E-1}
is of the same form as the
 conjectured one \eqref{nextorder},
 up to a  {\it positive} correction in the constant of order $\lambda$.
This constant $C$ and the constant
in the error term in \eqref{E-1} depends on the details of
the interaction potential, in particular our proof uses that $\hat V$ is
a soft potential.

The trial state in this paper consists only of condensate particles
and of non-condensate particle pairs with momenta
$k, -k$, reminiscent of the original idea of Bogoliubov.
 In our computation, however, the interactions between non-condensate
particle pairs are also relevant. Our trial state does not have a
fixed number of particles, but it is a state in the Fock space
with expected number of particles equal to $N$. It is
similar to the trial states used by Solovej \cite{Sol}
to give  rigorous upper bounds to the ground state energies
of the one and two-component
charged Bose gases in the high density limit.

Variational trial states with particle pairs have been  used earlier in the
context of the low density Bose gas by  Girardeau and Arnowitt \cite{GA}.
Their state, however, had a fixed number of particles which slightly
 complicated the calculation (the details are available only
in the unpublished Ph.D. dissertation of Girardeau).
 The variational formula we obtain is nevertheless
the same as theirs up to lower order terms due to the choice
of a different ensemble. However, in \cite{GA},
the solution of the minimization problem was given only implicitly
as a solution to a nonlinear  integral  equation and thus the energy was not
evaluated explicitly.  In our work, we identify the presumed main terms
from the calculations of each individual terms in the energy.
This enables us to find the minimizer for the main terms of the energy.
By choosing the minimizer of the main part
as our trial state, we {\it a-posteriori} justify
that the neglected terms are indeed of lower order
and thus giving a rigorous upper bound on the energy.
We believe that the difference between the energy of our
state and that of the true minimizer of the full functional
is of lower order.

\section{Setup}

We work in a finite box $\Lambda=\Lambda_L= [0,L]^3\subset \bR^3$
with periodic boundary conditions. Its dual space is
$\Lambda^*:= (\frac{2\pi}{L} \bZ)^3$.
For a continuous function $F$ on $\bR^3$, we have
\begin{equation*}
\begin{split}
\lim_{L \to \infty}   \frac{1}{L^3} \sum_{p\in\Lambda^*} F(p) = \; &
\lim_{L \to \infty}  \frac{1}{|\Lambda_L|} \sum_{p\in\Lambda^*} F(p)
 \\ = \; & \int_{\bR^3}
\frac{\rd^3 p}{(2\pi)^3} F(p)\; .
\end{split}
\end{equation*}
The Fourier
transform of an arbitrary function $f(x)$ on $\Lambda$ is defined as
$$
    \wh f_p = \int_\Lambda e^{-ip\cdot x} f(x) \rd x, \qquad
    f(x) = \frac{1}{|\Lambda|} \sum_{p\in \Lambda^*} e^{ip\cdot x} \wh f_p \; .
$$
Note that the Fourier transform depends on $\Lambda$, a fact that
is omitted from the notation. In most cases we will take Fourier
transforms of sufficiently decaying functions, so that
 their $\Lambda$ dependence
 is negligible in the limit $L\to\infty$.
Since  $V(x)$ is real and symmetric, we have that $\wh V_u$ is real and
$$
   {\widehat {V}}_u={\widehat V}_{-u} \; .
$$
We also have
$$
    \frac{1}{|\Lambda|} \sum_{p\in \Lambda^*} e^{ip\cdot x} = \delta_{\bR^3}(x),
\qquad \int_\Lambda e^{ip\cdot x}\rd x = \delta_{\Lambda^*}(p)
$$
where $\delta_{\bR^3}(x)$ is the usual continuum delta function and
$\delta_{\Lambda^*}$ is the lattice delta function, i.e.
$\delta_{\Lambda^*}(p)=
|\Lambda|= L^3$ if $p=0$, and $\delta_{\Lambda^*}(p)=0$ if $p\in\Lambda^*
\setminus \{ 0\}$.
We will neglect the subscripts in the delta functions, the argument
indicates whether it
is the momentum or position space delta function.
We also simplify the notation
$$
\sum_p:=   \sum_{p\in \Lambda^*}
$$
i.e. momentum summation is always on the whole $\Lambda^*$.

Notice that the choice of the $\delta_{\Lambda^*}$ ensures
that in the $L\to\infty$ limit, this delta function converges to
the usual continuum delta function $\delta(p)$ in momentum space
with respect to the measure $\rd^3 p/(2\pi)^3$:
\begin{equation*}
\begin{split}
      \lim_{L\to\infty}
      \frac{1}{|\Lambda|}\sum_{p\in\Lambda^*}& F(p) \delta_{\Lambda^*}(p-q)
      \\ &= \int_{\bR^3} \frac{\rd^3p}{(2\pi)^3} F(p)\delta(p-q)
      = F(q)
\end{split}
\end{equation*}
(where $\delta(p)$ is defined by the last formula).

\bigskip

Using the formalism of second quantization, we work on the bosonic Fock
space $\cF=\bigoplus_{n=0}^\infty \cH^{\otimes_s n}$
built upon the single particle Hilbert space
$\cH=\ell^2(\Lambda^*)$, where $\cH^{\otimes_s n}$ denotes the symmetric
tensor product of $n$ copies of $\cH$. The vacuum is denoted by $|0\rangle$.
We consider  bosonic annihilation and creation operators,
$\tilde  a_k , \tilde a^{+}_k$, for any $k\in\Lambda^*$,
with the usual canonical commutation relations (CCR):
$$
     [\tilde a_p,  \tilde   a^{+}_q] =
\tilde  a_p \tilde  a^{+}_q - \tilde   a^{+}_q \tilde   a_p = \delta(p-q)
    = \left\{
\begin{array}{ll}
 L^d & \mbox{ if } p=q \\
0 & \mbox{ otherwise.}
\end{array}
\right.
$$
The Hamiltonian of the system is given by
$$
     H = \frac{1}{|\Lambda|}
\sum_p p^2  \tilde  a^{+}_p \tilde  a_p + \frac{1}{2|\Lambda|^3} \sum_{p,q,u}
    {\widehat V}_u \tilde   a^{+}_p \tilde   a^{+}_q \tilde  a_{p-u} \tilde  a_{q+u} \; .
$$
where the first term is the kinetic energy,  the second term is
the interaction energy of the particles in appropriate physical units.
It is more convenient to redefine the bosonic operators as
$$
   a_k =  \frac{1}{\sqrt{|\Lambda|}} \; \tilde a_k,
\qquad \cre_k =  \frac{1}{\sqrt{|\Lambda|}}\; \tilde  a^{+}_k,
$$
i.e., from now on we assume that
$$
     [\ann_p, \cre_q] = \ann_p \cre_q -\cre_q\ann_p
    = \left\{
\begin{array}{ll}
 1 & \mbox{ if } p=q \\
0 & \mbox{ otherwise.}
\end{array}
\right.
$$
Thus the Hamiltonian is given by
\be
     H =
\sum_p p^2 \cre_p\ann_p + \frac{1}{2|\Lambda|} \sum_{p,q,u}
    {\widehat V}_u \cre_p\cre_q\ann_{p-u}\ann_{q+u} \;.
\label{ham}
\ee

\section{The trial state}

Let $c_k$ denote a family of complex numbers parametrized by
 $k\in \Lambda^*\setminus\{ 0\}$
with the property that $|c_k|<1$ and $c_k=c_{-k}$.
We define  a state
\be
   \Psi: = e^{\frac{1}{2}\sum_{k\neq0}
c_k \cre_k\cre_{-k}+\sqrt{N_0}\cre_0}|0\rangle\;,
\label{state}
\ee
where $N_0$ is a positive real
number. The parameters $c_k$ and $N_0$ will be fixed later on.

Fix a small positive number $\varrho$ which will be the density of the system
and  let $N$ denote
$$
N:=\varrho {|\Lambda|}\; .
$$
Define the expectation w.r.t. $\Psi$ by
$$
 \langle A   \rangle_{\Psi}  = \frac{\langle\Psi,  A\Psi\rangle}{
\langle \Psi , \Psi\rangle}\; ,
$$
where $\langle \cdot, \cdot\rangle$ denote the standard $L^{2}$ inner product.
We shall fix the parameters in $\Psi$ so that the expected number of particles is given by $N$
\be
    N = \Big\langle \sum_{m\in \Lambda^{*}} \cre_{m}\ann_{m}\Big\rangle_{\Psi}\; .
\label{number}
\ee
Let $E = \langle  H   \rangle_{\Psi}$ be the energy.

Let $1-w$ be the zero energy scattering solution to the potential $V$
\be
   -\Delta(1-w) + \frac{1}{2} V(1-w)=0
\label{scateq}
\ee
on $\bR^3$
with $0\leq w<1$ and $w(x)\to 0$ as $|x|\to \infty$.
Then the scattering length is defined by
\be
\begin{split}
  a: = \; &\frac{1}{8\pi}\int_{\bR^3} V(x)(1-w(x))\rd x \\ = \; &
  \frac{1}{8\pi}\lim_{L\to\infty} \int_{\Lambda_L} V(x)(1-w(x))\rd x\; .
\label{def:a}
\end{split}
\ee
It is well-known that
\be
8 \pi a < \int_{\bR^3} V(x) \rd x  =
\lim_{L\to\infty} \int_{\Lambda} V(x)\rd x = \lim_{L\to\infty} \wh V_0 \, ,
\label{8pia}
\ee
where, in the last step, we recall that the definition of the Fourier transform depends on $L$.

Define  the number $h$ by
\be\label{hdef}
h= \frac {{\widehat V}_{0} } {8 \pi a} - 1 \, ,
\ee
from \eqref{8pia} it follows that $h>0$ if $L$ is sufficiently large.
Recall that $V = \lambda \tilde V$ with $\tilde V$ being fixed.
The scattering length $a$ can be computed via the Born series
for small $\lambda$.  In particular, we will show in Lemma
\ref{lemma:scatt} that $h$ is of order $\lambda$
\be
     h = O(\lambda) \; .
\label{hlambda}
\ee
Define the function
\be\begin{split}\label{def:Phi}
  \Phi(h) &:= \int_{0}^{\infty} \rd y \; y^{1/2} \\
 &\times \Big(\sqrt{(y+2h)(y+2+2h)}-(y+1+2h) +\frac{1}{2y}\Big)
 \; .
\end{split}
\ee
One can check that this integral is convergent for $h\ge 0$.
Our main result is the following theorem.

\begin{theorem}\label{mainth}
Let $ \tilde { V}(x) \ge 0$, $\tilde V\not\equiv0$,
  be a non-negative radially symmetric  smooth function with a decay
$\tilde V(x)\leq C(1+|x|)^{-3-\delta}$ for some $\delta>0$, 
 and set $ { V}(x) = \lambda \tilde   V(x)$.
 Then for $\lambda$ small enough, we have, in the limit
 $\varrho \to 0$, the following estimate
\be
 E= 4\pi\varrho Na +   Q    + O(N\varrho^2|\log \varrho|)
\label{eq:E0}
\ee
for the energy of the trial state \eqref{state} with an appropriate choice
of $c_k$ and $N_0$,
under the constraint \eqref{number}.
Here  $ Q = Q(h)$ is given by
$$
 Q (h)
 = 4\pi a N\varrho \Bigg[ \sqrt{\frac{32}{\pi}} \Phi(h) (a^3\varrho)^{1/2}
\Bigg]
$$
 and the constant in the error term in \eqref{eq:E0} depends on $\lambda$ and $\tilde V$.
\end{theorem}
The assumptions on $\tilde {V}$ can certainly be relaxed but
we do not aim at identifying the optimal conditions.

A direct calculation gives
$$
  \Phi(0)= \frac{\sqrt{512}}{15}\; ,
$$
thus at $h=0$ we obtain
$$
  Q (0)
 = 4\pi a N\varrho \frac{128}{15\sqrt{\pi}}(a^3\varrho)^{1/2} \; .
$$
Moreover,
a simple calculation also shows that
$$
   0< \Phi'(0)= \int_0^\infty \frac{2y^{1/2}\rd y}{(y+1)\sqrt{y(y+2)}} <\infty,
$$
thus infinitesimally $\Phi(0)< \Phi(h)$ if $0<h\ll 1$ and
$$
   Q(h) = Q(0) + O(h ) = Q(0) + O(\lambda)
$$
by \eqref{hlambda}.
In fact, it is also easy
to see that
$$
   \Phi(0)< \Phi(h)
$$
holds for any $h>0$.
Thus  our trial state delivers a second order
term with an explicit constant that is bigger than the Lee-Yang
prediction \cite{LYang, LHY}  by a factor $(1+O(\lambda))$
for small coupling constant $\lambda$.

\section{Computation of the Energy}

We start the proof of the main theorem by the following Lemma.
We first define a few quantities:
\be\begin{split}\label{omega2}
\Omega_{2}= \; &- \sum_{p\ne 0} \frac{({\widehat V}_p+{\widehat V}_0)}{|\Lambda|}
\left (\sum_{m\neq 0} \frac{|c_m|^2}{1-|c_m|^2} \right )  \frac{|c_p|^2}{1-|c_p|^2} \\ &+
 \frac{{\widehat V}_0}{2|\Lambda|} \Big( \sum_{m\ne 0}
 \frac{|c_m|^2}{1-|c_m|^2}\Big)^2
\end{split}\ee
\begin{equation}\begin{split}\label{omega3}
\Omega_{4}  = \;& \sum_{p\ne 0} \Bigg[  \frac{1}{2|\Lambda|}
\sum_{r\neq 0, \pm p} ({\widehat V}_0 +{\widehat V}_{p-r}) \frac{|c_p|^2|c_r|^2}{(1-|c_p|^2)
  (1-|c_r|^2)}  \\
  &+  \frac{1}{2|\Lambda|}
\frac{{\widehat V}_0|c_p|^2(1+3 |c_p|^2)+
{\widehat V}_{2p}|c_p|^2(1+|c_p|^2) }{(1-|c_p|^2)^2}\Bigg] \;.
\end{split}
\end{equation}

\begin{lemma}\label{lm:EM}
The energy $E=\langle H \rangle_{\Psi}$ of the state \eqref{state} under the constraint
\eqref{number}
 is given by $E=E_{M}+ \Omega_{2}+ \Omega_{4}$,  where
\begin{align}\label{EE}
  E_{M} := & \frac{1}{2|\Lambda|}{\widehat V}_0N^2 \\ &+
\sum_{p\ne 0} \Bigg[(p^2+\varrho {\widehat V}_p) \frac{|c_p|^2}{1-|c_p|^2}
 +  \varrho {\widehat V}_p\frac{\mbox{Re} \, c_p}{1-|c_p|^2}   \nonumber \\
&  + \frac{1}{2|\Lambda|}\sum_{r\ne 0,\pm p}  {\widehat V}_{p-r}
\frac{\bar c_pc_r}{(1-|c_p|^2)
  (1-|c_r|^2)} \nonumber  \\ &- \frac{{\widehat V}_p}{|\Lambda|}
\sum_{m\ne 0} \frac{|c_m|^2}{1-|c_m|^2}
 \frac{\mbox{Re} \, c_p}{1-|c_p|^2}  \Bigg ] \; .\nonumber
\end{align}
\end{lemma}

Remark: We shall see later on that the first four terms
in the main term, $E_{M}$, are of order $N \varrho$,
the fifth one is  of order $N \varrho^{3/2}$. Each term
in the error terms $\Omega_{2}$ and $\Omega_4$
is at most of order $N\varrho^{2}|\log\varrho|$ except the last term in
$\Omega_4$ which is
non-extensive.

\bigskip

\noindent {\bf Proof.} We collect a few elementary  facts from direct
calculations. Similar formulas in an abstract setup applicable
in general, not only for the translation invariant case, were
presented in \cite{Sol}.
\begin{align}
   \langle 0 | e^{\bar c_k \ann_k\ann_{-k}} e^{c_k\cre_k\cre_{-k}}|0\rangle
  &= \frac{1}{1-|c_k|^2} \quad \mbox{for}\;\; k\neq 0\;,  \nonumber\\
   \langle \Psi , \Psi\rangle
&= e^{N_{0}} \prod_{k\neq0} \frac{1}{\big(1-|c_k|^2\big)^{1/2}}\; ,
\nonumber\\
    \lan \cre_0\ann_0\ra_{\Psi} & = N_0\;, \nonumber\\
\lan \cre_0\cre_0\ann_0\ann_0\ra_{\Psi} & = N_0^2, \nonumber
\end{align}
and  for $m\neq 0$,
\begin{align}
  \lan \cre_m\ann_m\ra_\Psi & = \frac{|c_m|^2}{1-|c_m|^2} \nonumber \\
  \overline{\lan \cre_m \cre_{-m} \ra}_{\Psi}
 = \lan \ann_m \ann_{-m} \ra_{\Psi} & = \frac{c_m}{1-|c_m|^2} \nonumber \\
  \lan \cre_m\cre_m\ann_{-m}\ann_{-m}\ra_\Psi &= 0\nonumber\\
  \lan \cre_m\cre_{-m}\ann_m\ann_{-m}\ra_\Psi &=
 \frac{|c_m|^2(1+|c_m|^2)}{(1-|c_m|^2)^2} \nonumber\\
  \lan \cre_m\cre_m\ann_m\ann_m\ra_\Psi & =
  \frac{2|c_m|^4}{(1-|c_m|^2)^2} \nonumber \;.
\end{align}
Moreover,
\be
   \lan A\ann_m B\ra_\Psi = 0 , \qquad m\neq 0\; ,
\label{AB}
\ee
where $A$ and $B$ are any operator not containing $\cre_{m}$ or $\ann_{-m}$.
In fact, for an appropriate operator $C$ commuting with
$\ann_m, \ann_{-m}, \cre_m, \cre_{-m}$, we have
\begin{equation*}
\begin{split}
\lan A &\, \ann_m B \ra_{\Psi} \\ = \; &\lan 0 | \, C^* e^{\bar{c}_m \ann_m \ann_{-m}} A
 \,\ann_m B e^{c_m \cre_m \cre_{-m}} C |0 \rangle \\ = \; & \sum_{n,k \geq 0}
\frac{\bar{c}_m^n c_m^k}{n! k!} \, \lan 0 | \, C^* \, (\ann_m \ann_{-m})^n A \,
\ann_m B (\cre_m \cre_{-m})^k \, C  | 0 \ra \\ = \; & \sum_{n,k \geq 0}
\frac{\bar{c}_m^n c_m^k}{n! k!} \\ &\hspace{.5cm}
 \times \lan 0 | (\ann_{-m})^n (\cre_{-m})^k \, C^* A B C \, (\ann_m)^{n+1} (\cre_m)^k  | 0 \ra\,.
\end{split}
\end{equation*}
Equation (\ref{AB}) follows because, on the one hand, $(\ann_m)^{n+1} (\cre_m)^k |0 \ra = 0$
 if $n+1 >k$ and, on the other hand $\lan 0| (\ann_{-m})^n (\cre_{-m})^k = 0$ if $k>n$.


The total particle number is computed as
\begin{equation*}\begin{split}
 \big\lan \sum_{m} \cre_m\ann_m\big\ra_\Psi = \; & \lan \cre_0\ann_0\ra_\Psi +
   \sum_{m\neq 0} \lan \cre_m\ann_m\ra_\Psi \\ = \; &N_0 + \sum_{m\ne0}
\frac{|c_m|^2}{1-|c_m|^2}\; ,
\end{split}\end{equation*}
so the constraint \eqref{number} is equivalent to
\be
     N = N_{0}+ \sum_{m\ne 0}\frac{|c_{m}|^{2}}{1-|c_{m}|^{2}} \; .
\label{tot}
\ee
We shall see after  \eqref{orderc} that with  our choice of parameters
$c_{m}$ and $N_{0}$ we have
\be\label{n1}
\frac {N-N_{0}} N \sim  C\lambda^{3/2}\varrho^{1/2} +  O(\lambda^2\varrho);
\ee
with a positive
constant $C$ that depends only on the unscaled potential $\hat V_0$.
In particular, the depletion rate of the condensate
is of order $\lambda^{3/2}\varrho^{1/2}$.

We classify the interaction terms in the Hamiltonian \eqref{ham}
according to their number of zero momentum operators, $\cre_0$ or
$\ann_0$. It will turn out that only even number of zero momentum
operators give non-zero contribution. We
will thus write
$$
\Big\lan \frac{1}{2|\Lambda|} \sum_{p,q,u}
    {\widehat V}_u \cre_p\cre_q\ann_{p-u}\ann_{q+u}\Big\ra_\Psi
 = E_0 + E_2 + E_4\;,
$$
where $E_k$, $k=0,2,4$, defined below, denote the contributions
of terms with exactly
$k$ zero momentum operators.

\medskip

{\it Case 1.} All four operators are with zero momentum, i.e.
 $p=q=u=0$, and the contribution of this part is
\be
E_{0}= \frac{1}{2|\Lambda|}{\widehat V}_0 N_0^2 \; .
\ee

{\it Case 2.} By momentum conservation in the interaction term,
it is impossible to have exactly three zero momentum operators.
The terms containing  exactly two zero momentum operators are
\be\begin{split}\label{2}
   \frac{1}{2|\Lambda|} \sum_{u\ne 0}
 \Big( & \,{\widehat V}_u \cre_0\cre_0\ann_u\ann_{-u}
+ {\widehat V}_u \cre_u\cre_{-u}\ann_0\ann_0 \\ &+ 2({\widehat V}_u+{\widehat V}_0)\cre_u\cre_0\ann_0\ann_u
 \Big) \; .\end{split}
\ee
The contribution of this term to the potential energy of $\Psi$ is
\be\label{2.1}
E_{2}= \sum_{p\ne 0} \Bigg[\frac{N_0{\widehat V}_p}{2|\Lambda|} \frac{c_p+\bar c_p}{1-|c_p|^2}
 + \frac{2({\widehat V}_p+{\widehat V}_0)N_0}{2|\Lambda|} \frac{|c_p|^2}{1-|c_p|^2} \Bigg ] \; .
 \ee

  \bigskip

Suppose that
among the four momenta, $p$, $q$, $p-u$, $q+u$, exactly
one is zero, say $p-u$ (other cases are analogous).
Then the remaining three operators are
$\cre_p\cre_q\ann_{p+q}$. Since $p,q,p+q$ are nonzero, either
$p,q,p+q, -p,-q,-(p+q)$ are all different, or
$p=q$, in which case $\ann_{p+q}$ stands alone without
any other operator $\cre_{\pm(p+q)}$ or $\ann_{\pm(p+q)}$.
{F}rom \eqref{AB}, the expectation of this term with respect to $\Psi$ vanishes. This proves that
there is no contribution for the case of exactly one zero momentum operator.

\bigskip

{\it Case 3.} No zero momentum operator is present in the
interaction term in \eqref{ham}. Let $r:=p-u$, $s:=q+u$.
 Based upon \eqref{AB}, only the following cases yield a non-zero
 contribution:

\begin{itemize}
\item $p=-q$ and $r\not\in\{ p,-p\}$, but since $r$ must be $\pm s$
and we have momentum conservation, $r=-s$. The energy contribution is
the main term in this case:
\begin{equation*}
\begin{split}
\tilde E_{4} : = \;&   \frac{1}{2|\Lambda|} \sum_{p\neq0} \sum_{r \neq0,\pm p} {\widehat V}_{p-r}
\lan \cre_p\cre_{-p}\ann_r\ann_{-r} \ra_\Psi  \\ = \; &
 \frac{1}{2|\Lambda|} \sum_{p \neq 0} \sum_{r \neq 0, \pm p} {\widehat V}_{p-r}
\frac{\bar c_p c_r}{(1-|c_p|^2)(1-|c_r|^2)} \; ;
\end{split}
\end{equation*}

\item $p=-q=r=-s$
$$
   \frac{1}{2|\Lambda|} {\widehat V}_0\sum_{p\neq0}
\lan \cre_p\cre_{-p}\ann_p\ann_{-p} \ra_\Psi  =
 \frac{{\widehat V}_0}{2|\Lambda|}\sum_{p\neq0}
\frac{|c_p|^2(1+|c_p|^2)}{(1-|c_p|^2)^2} \; ;
$$

\item $p=-q=-r=s$
\begin{equation*}\begin{split}
   \frac{1}{2|\Lambda|} \sum_{p\neq0} {\widehat V}_{2p}
&\lan \cre_p\cre_{-p}\ann_{-p}\ann_{p} \ra_\Psi  \\ &=
 \frac{1}{2|\Lambda|}\sum_{p\neq0} {\widehat V}_{2p}
\frac{|c_p|^2(1+|c_p|^2)}{(1-|c_p|^2)^2} \; ;
\end{split}
\end{equation*}

\item $p=q$, then $r+s=2p$ and $r=\pm s$ implies $r=s=p$ and
we have
$$
   \frac{1}{2|\Lambda|}  \sum_{p\neq0} {\widehat V}_{0}
\lan \cre_p\cre_{p}\ann_{p}\ann_{p} \ra_\Psi  =
 \frac{{\widehat V}_0}{2|\Lambda|}\sum_{p\neq0}
\frac{2 |c_p|^4}{(1-|c_p|^2)^2} \; ;
$$

\item $p=r$, $q=s$ and $p=s$, $q=r$ but $p\neq \pm q$:
\begin{equation*}
\begin{split}
  \frac{1}{2|\Lambda|}  &\sum_{p\neq 0} \sum_{q\neq 0, \pm p} {\widehat V}_{0}
\lan \cre_p\cre_{q}\ann_{p}\ann_{q} \ra_\Psi
\\ &+  \frac{1}{2|\Lambda|}  \sum_{p\neq0} \sum_{q \neq 0, \pm p} {\widehat V}_{p-q}
\lan \cre_p\cre_{q}\ann_{q}\ann_{p} \ra_\Psi \\
= \; &\frac{1}{2|\Lambda|}\sum_{p\neq 0} \sum_{ q\neq 0, \pm p}  ({\widehat V}_0+{\widehat V}_{p-q})
\frac{|c_p|^2|c_q|^2}{(1-|c_p|^2)(1-|c_q|^2)} \; .
\end{split}
\end{equation*}
\end{itemize}
Collecting all these terms, the contribution of the case 3 to the potential energy is
\be\label{3}
E_{4} = \tilde E_{4}+ \Omega_{4}\; ,
\ee
where $\Omega_4$ was defined in  \eqref{omega3}.

We now combine the contribution to the potential energy from case 1 and case 2 and
we use the relation between $N$ and $N_0$ given by \eqref{tot}:
\begin{align}\label{case12}
E_{0} &+ E_{2} = \\=& \frac{1}{2|\Lambda|}{\widehat V}_0N^2_0 \nonumber \\ &+
\sum_{p\ne 0} \Bigg[
\frac{N_0{\widehat V}_p}{|\Lambda|} \frac{\mbox{Re} \, c_p }{1-|c_p|^2}
 + \frac{2({\widehat V}_p+{\widehat V}_0)N_0}{2|\Lambda|} \frac{|c_p|^2}{1-|c_p|^2}\Bigg]
 \nonumber \\
 = & \frac{1}{2|\Lambda|}{\widehat V}_0N^2 -
\frac{N}{|\Lambda|}{\widehat V}_0 \sum_{m\neq 0} \frac{|c_m|^2}{1-|c_m|^2}
\nonumber \\ & +  \frac{{\widehat V}_0}{2|\Lambda|} \Big( \sum_{m\ne 0}
 \frac{|c_m|^2}{1-|c_m|^2}\Big)^2  \nonumber \\
& +\sum_{p\ne 0} \Bigg[
 \frac{N {\widehat V}_p}{|\Lambda|} \frac{\mbox{Re} \, c_p}{1-|c_p|^2}
 + \frac{({\widehat V}_p+{\widehat V}_0)N}{|\Lambda|} \frac{|c_p|^2}{1-|c_p|^2}\Bigg] \nonumber \\
 &- \sum_{p\ne 0} \Bigg[
 \frac{ {\widehat V}_p}{|\Lambda|}  \left (\sum_{m\neq 0} \frac{|c_m|^2}{1-|c_m|^2} \right )
\frac{\mbox{Re} \,c_p}{1-|c_p|^2}
\nonumber \\ &\hspace{.5cm}+ \frac{({\widehat V}_p+{\widehat V}_0)}{|\Lambda|} \left (\sum_{m\neq 0}
\frac{|c_m|^2}{1-|c_m|^2} \right )  \frac{|c_p|^2}{1-|c_p|^2}\Bigg] \; .\nonumber
\end{align}
Notice that there are two  terms of the form $-  \frac{N}{|\Lambda|}{\widehat V}_0
 \sum_{m\neq 0} \frac{|c_m|^2}{1-|c_m|^2} $
which cancel each other. So we can rewrite \eqref{case12} as
\be
E_{0}+ E_{2} = \tilde E_{0} + \tilde E_{2} + \Omega_{2} \; ,
\ee
where
\be\label{E0}
\tilde E_{0}= \frac{1}{2|\Lambda|}{\widehat V}_0N^2 \; ,
\ee
\be\label{case12f}
\begin{split}
 \tilde E_{2} =
 \sum_{p\ne 0} \Bigg[  & \varrho  {\widehat V}_p  \frac{|c_p|^2}{1-|c_p|^2} +
 \varrho  {\widehat V}_p  \frac{\mbox{Re}\, c_p}{1-|c_p|^2}
\\ &- \frac{ {\widehat V}_p}{|\Lambda|}  \left (\sum_{m\neq 0} \frac{|c_m|^2}{1-|c_m|^2} \right )
\frac{\mbox{Re}\, c_p}{1-|c_p|^2} \Bigg]
\end{split} \ee
and $\Omega_2$ is given in \eqref{omega2}.
The first term in \eqref{case12f}, when combined with the
kinetic energy contribution
\[
\sum_{p\ne 0} p^2 \frac{|c_p|^2}{1-|c_p|^2} \; ,
\]
is the second term in \eqref{EE}.  The remaining main terms in \eqref{EE}
come from the rest of $\tilde E_{2}$, $\tilde E_{0}$ and $\tilde E_{4}$, i.e.
\be
  E_M = \sum_{p\ne 0} p^2 \frac{|c_p|^2}{1-|c_p|^2} + \tilde E_0 + \tilde E_2 +\tilde E_4 \, .
\label{EM}
\ee
This completes the proof of Lemma \ref{lm:EM}. $\;\;\Box$

\bigskip
Notice that the main terms in the potential energy come from the following channels:
\begin{align}\label{main-ch}
  \frac{1}{2|\Lambda|}  \cre_{0} &\cre_{0}\ann_{0}\ann_{0} +
   \frac{1}{2|\Lambda|} \sum_{u\ne 0}
 \Big( {\widehat V}_u \cre_0\cre_0\ann_u\ann_{-u}
\nonumber \\ &+ {\widehat V}_u \cre_u\cre_{-u}\ann_0\ann_0 +
2({\widehat V}_u+{\widehat V}_0)\cre_u\cre_0\ann_0\ann_u
 \Big) \nonumber \\
 &
 +  \frac{1}{2|\Lambda|} \sum_{p\neq0} \sum_{r \neq 0,\pm p} {\widehat V}_{p-r}
 \cre_p\cre_{-p}\ann_r\ann_{-r}   \; .
\end{align}
The main energy contribution from these terms are all of order $N \varrho$.
In the last term, the interaction between two large momenta ($|p|, |r|\sim 1$) pairs
contribute with order $N\varrho$. The order
$N \varrho^{3/2}$ term comes partly from substituting $N_0$, the expected value of $\cre_0\ann_0$,
with $N$ (using that  $N-N_{0}\sim CN \varrho^{1/2}$) and  partly from
the interaction between a low momentum pair, $|p|\ll 1$, and a large momentum pair,
$|r|\sim 1$.

\section{The one-particle scattering problem}

Recall that $1-w$ was the solution to the zero energy scattering equation \eqref{scateq}
and $\wh w_p$ denotes the Fourier transform of $w$.
If $V$ is smooth and it decays sufficiently fast at infinity,  then $w(x)$ is smooth and
$w(x) \leq C|x|^{-1}$ for large $|x|$. Its Fourier transform on
 $\bR^3$, $\int_{\bR^3} e^{-ip\cdot x}w(x) \rd x$,
has a  $|p|^{-2}$ singularity at the origin. The lattice Fourier transform
satisfies the regularized bound
$$
   |\wh w_p| \leq \wh w_0  =\int_\Lambda w(x) \rd x \leq CL^2\; , \qquad  p\in \Lambda^*\; .
$$
This bound guarantees that for any function $\varphi\in L^1(\bR^3)$, the lattice Fourier
transform of $\varphi w$ can be computed as
\begin{equation*}
\begin{split}
  \widehat{(\varphi w)}_p &=  (\wh\varphi\ast\wh w)_p=
 \frac{1}{|\Lambda|}\sum_{r\in\Lambda^*} \wh \varphi_{p-r} \wh w_r \\ &
  =\frac{1}{|\Lambda|}\sum_{r\neq 0} \wh \varphi_{p-r} \wh w_r + O\Big(\frac{1}{L}\Big) \; .
\end{split}
\end{equation*}
Thus, modulo an error that is negligible in the thermodynamic limit, we can restrict
the momentum summations involving $\wh w_r$ to $r\neq 0$.

{F}rom the definition of the scattering length \eqref{def:a}, we have
\be
    8\pi a = {\widehat V}_0 - \int {\widehat V}_p \wh w_p  \frac{\rd^3p}{(2\pi)^3}
    = {\widehat V}_0 - \frac{1}{|\Lambda|}\sum_{p\neq 0} {\widehat V}_p\wh w_p + O\Big(\frac{1}{L}\Big)\; .
\label{av}
\ee
{F}rom the scattering equation we get
\be
  -p^2 \wh w_p+ \frac{1}{2}{\widehat V}_p -\frac{1}{2|\Lambda|} \sum_{r\neq 0}
 {\widehat V}_{p-r} \wh w_r= O\Big(\frac{1}{L}\Big)
\qquad \forall p \neq 0\, .
\label{scateq1}
\ee
Define
\be
    f(x):= V(x) w(x), \qquad g(x): =  V(x) - f(x)\, ,
\label{gdef}
\ee
then in Fourier space we have
$$
 \wh f_p = ({\widehat V}\ast \wh w)_p = \frac{1}{|\Lambda|}
 \sum_{r\neq0} {\widehat V}_{p-r} \wh w_r +  O\Big(\frac{1}{L}\Big) \; .
$$
In particular, from \eqref{av}
\be
    8\pi a =\wh g_0 +  O\Big(\frac{1}{L}\Big) \; .
\label{ag}
\ee
In the sequel we will not
carry the negligible error term $O(1/L)$ in the formulas.

\begin{lemma}\label{lemma:scatt} Let  $\tilde V(x) \ge 0$, $\tilde V \not\equiv 0$,
be a radially symmetric smooth function with a sufficiently fast decay as $|x| \to \infty$
and let $V=\lambda\tilde V$.
Then, for a sufficiently small $\lambda$,
\be
  {\widehat V}_p, \wh f_p, \wh g_p  \quad \mbox{are real and have a fast  decay as $|p|\to \infty$,}
\label{cond}
\ee
\be\label{cond2}
  0< \wh f_0, \wh g_0< {\widehat V}_0 \; .
\ee
Moreover
\be
        0<  \frac{\wh V_0}{8\pi a}-1= \frac{\wh f_0}{\wh g_0} = O(\lambda)\; .
\label{f0g0}
\ee
Furthermore, $\wh f_{p}, \wh g_{p}, {\widehat V}_p$ are uniformly Lipschitz continuous, i.e.,
\be\label{lip}
\begin{split}
\sup_p\Big(   |\wh g_p-\wh g_{p-r}|+
  |{\widehat V}_p-{\widehat V}_{p-r}| \Big)
& \leq C\lambda|r| \; ;\\
\sup_p |\wh f_p-\wh f_{p-r}|& \leq C\lambda^2|r|\; ,
\end{split}
\ee
with a constant $C$ depending only on $\tilde V$. All statements hold uniformly
in the thermodynamic limit, i.e. for all $L$ sufficiently large.
\end{lemma}

\noindent
{\bf Proof.} The reality of $\wh V_p, \wh f_p, \wh g_p$ follows from the radial symmetry.
{F}rom the scattering equation \eqref{scateq1}
\be\label{fg}
   -2p^2\wh w_p +{\widehat V}_p - \wh f_p =0, \qquad  \wh g_p= 2p^2\wh w_p
 \qquad \forall p\neq 0 \; .
\ee
By iteration, we obtain the Born series
for the scattering wave function ($p\neq 0$)
\be
   \wh w_p= \frac{{\widehat V}_p}{2p^2} - \frac{1}{2p^2}\sum_{r\neq 0}
 \frac{{\widehat V}_{p-r}{\widehat V}_r}{2r^2}
+ \frac{1}{2p^2}\sum_{r,u\neq 0} \frac{{\widehat V}_{p-r}{\widehat V}_{r-u}{\widehat V}_u}{4r^2u^2}
-\ldots
\label{born}
\ee
It is easy to see from the expansion \eqref{born}
 that \eqref{cond} is satisfied if ${\widehat V}_p$ is sufficiently
small and decaying. Furthermore, $\wh f_0 = |\Lambda|^{-1}
\sum_{p\neq 0} \wh V_p \wh w_p = O(\lambda^2)$,
while $\wh V_0 = c\lambda$ with a positive constant $c=\int \tilde V$, thus
$\wh g_0 = \wh V_0-\wh f_0 \ge c\lambda/2$ if $\lambda$ is
sufficiently small and \eqref{f0g0} follows.
 The rest of the statements of the Lemma also easily follows
from  \eqref{born}. $\;\;\Box$

\section{The minimization}

{F}rom now on we assume that $c_p$ are real, it is most likely that complex choice does not
lower the energy of our trial state. We introduce new variables
$$
   e_p:= \frac{c_p}{1+c_p}, \qquad c_p= \frac{e_p}{1-e_p} \; .
$$
{F}rom $|c_p|<1$ we have $e_p\in (-\infty, \frac{1}{2})$.
We also have the relations
\be
   \frac{c_p^2}{1-c_p^2} = \frac{e_p^2}{1-2e_p},
 \qquad  \frac{c_p}{1-c_p^2} = \frac{e_p(1-e_p)}{1-2e_p}.
\label{rel}
\ee

We shall choose $e_{p}$ via the following Lemma. This choice will become clear later on.
\begin{lemma}\label{le-ep} For any sufficiently small $\varrho\leq \varrho_0(\lambda)$
let  $-\infty <  e_p < 1/2$ be the minimizer of
\be\label{min}
 m_p := \min_{e_p}\Big[  p^2
  \frac{e_p^2}{1-2e_p} + \varrho {\widehat V}_p  \frac{e_p}{1-2e_p}
 -\varrho \wh f_p e_p\Big]\; .
\ee
Then
\be
    e_p = \frac{1}{2}\Big[1 -\Big(1+2 \frac{\varrho \wh g_p}{p^2+2\varrho \wh f_p}\Big)^{1/2}\Big]
\label{ep}
\ee
and the minimal value is given by
\be
  m_p= \frac{1}{2}\Big[ \sqrt{ (p^2+2\varrho \wh V_p)(p^2 + \varrho \wh f_p)} -
 \big( p^2 + \varrho(\wh V_p+ \wh f_p)\big)\Big] \;.
\label{minval}
\ee
\end{lemma}

\noindent
{\bf Proof.} Consider the minimization problem
$$
  \min_{e< 1/2} \Big[ a \frac{e^2}{1-2e} + b  \frac{e}{1-2e}
 -c e\Big] \; .
$$
where the parameters satisfy $a+2c>0$ and $a+2b>0$.
After differentiating in $e$, the minimizers satisfy the equation
$$
  \frac{2(a+2c)(e-e^2) +b-c}{(1-2e)^2} = 0 \; .
$$
Solving the quadratic equation, we get
\be
   e= \frac{1}{2}\Big[1\pm\Big(1+2 \frac{b-c}{a+2c}\Big)^{1/2}\Big]
\label{e}
\ee
and by the conditions on $a,b,c$ we have $1+2(b-c)/(a+2c)\ge 0$.
In our case, since $e< \frac{1}{2}$, only the negative sign can be correct.
With this choice, the minimum value is
\begin{equation*}\begin{split}
  \Big[ a \frac{e^2}{1-2e} + &b  \frac{e}{1-2e}
 - c e\Big] \\ &= \frac{1}{2}\Big[\sqrt{(a+2b)(a+2c)}-(a+b+c)\Big] \; .
\end{split}
\end{equation*}

In our application, the conditions $a+2c>0$, $a+2b>0$ are
equivalent to
\be
    p^2 +2\varrho \wh f_p >0, \qquad p^2 + 2\varrho \wh V_p>0
\label{inn}
\ee
and they  are always satisfied if $\varrho$ is sufficiently small.
In the regime $|p|\ge 4(\varrho \wh V_0)^{1/2}$, these inequalities
 follow from the bounds  $|\wh V_p|\leq V_0$ and $|\wh f_{p}|\leq \wh f_0 \leq \wh V_0$
(see \eqref{cond2}). For $|p|\leq 4(\varrho \wh V_0)^{1/2}$ we
have $p^2 + 2\varrho \wh f_p = p^2 + 2\varrho \wh f_0 + O(\varrho^{3/2})
\ge p^2 + \varrho \wh f_0 > 0 $
by  the Lipschitz continuity of $\wh f_p$ \eqref{lip} and the lower bound $\wh f_0 >0$.
The other inequality in \eqref{inn} is proven analogously. Actually, these
calculations also show that $p^2 +2\varrho \wh f_p$ and $p^2 + 2\varrho \wh V_p$
 have a positive lower bound
uniformly in $p$, if $\varrho$ is sufficiently small:
\be
   \inf_{p}  (p^2 +2\varrho \wh f_p) > 0, \qquad  \inf_{p} ( p^2 +2\varrho \wh V_p) > 0 \; .
\label{uniflow}
\ee
$\Box$

\vskip 1 cm

We now rewrite the error terms $\Omega_{2}+ \Omega_4$ in terms of $e_p$:
\begin{align}\label{E2}
\Omega_{2}&+ \Omega_4 \nonumber \\ = \; & \sum_{p\ne 0} \Bigg[
\frac{1}{2|\Lambda|}\sum_{r\neq 0, \pm p}
({\widehat V}_0+{\widehat V}_{p-r}) \frac{e_p^2e_r^2}{(1-2e_p)(1-2e_r)}\nonumber \\
 & \hspace{.5cm} +  \frac{1}{2|\Lambda|}\Big(\frac{e_p(1-e_p)}{1-2e_p}\Big)^2
\Big({\widehat V}_0
 \frac{1-2e_p+4e_p^2}{(1-e_p)^2} \nonumber \\ &\hspace{1cm} + {\widehat V}_{2p}
 \frac{1-2e_p+2e_p^2}{(1-e_p)^2}\Big)  \nonumber  \\
& \hspace{.5cm} - \frac{{\widehat V}_p+{\widehat V}_0}{|\Lambda|}
\Big(\sum_{r\ne 0} \frac{e_r^2}{1-2e_r}\Big)
 \frac{e_p^2}{1-2e_p}  \Bigg]
\nonumber \\ &+ \frac{{\widehat V}_0}{2|\Lambda|} \Big( \sum_{p\ne 0}
 \frac{e_p^2}{1-2e_p}\Big)^2 \; .
\end{align}
For the main term \eqref{EE}, we replace  ${\widehat V}_0$ with
$8\pi a+|\Lambda|^{-1} \sum_{p\ne 0} {\widehat V}_p \wh w_p $ in $\tilde E_0$
(see \eqref{E0} and \eqref{EM})
by using \eqref{av} at the expense of a term of order $1/L$ that is negligible
in the thermodynamic limit. Thus, neglecting this error term, we have
\begin{align}\label{E2.5}
  E_{M}= \; & 4\pi aN\varrho+ \frac{\varrho^2}{2}
 \sum_{p\ne 0} {\widehat V}_p \wh w_p  \nonumber \\ & +
\sum_{p\ne 0} \Bigg[(p^2+\varrho {\widehat V}_p) \frac{e_p^2}{1-2e_p}
+  \varrho {\widehat V}_p\frac{e_p(1-e_p)}{1-2e_p}  \nonumber \\
&\hspace{.5cm}    + \frac{1}{2|\Lambda|}\sum_{r\ne 0,\pm p}  {\widehat V}_{p-r}
\frac{e_p(1-e_p)e_r(1-e_r)}{ (1-2e_p)(1-2e_r)}  \nonumber \\
&\hspace{.5cm}- \frac{{\widehat V}_p}{|\Lambda|}
\Big(\sum_{r\ne 0} \frac{e_r^2}{1-2e_r}\Big)
 \frac{e_p(1-e_p)}{1-2e_p}  \Bigg ]  \; .
\end{align}

By using \eqref{cond2} and \eqref{lip} we have that
for a sufficiently small but fixed $\delta$ (depending on $V$),
$$
     \frac{{\widehat V}_0}{2}\leq {\widehat V}_p\leq {\widehat V}_0, \qquad
 \frac{{\widehat f}_0}{2}\leq {\widehat f}_p\leq {\widehat f}_0,\qquad
 \frac{{\widehat g}_0}{2}\leq {\widehat g}_p\leq {\widehat g}_0,
$$
for all $|p|\leq \delta$. In particular we have
\be
 \frac{\wh g_0}{4p^2} \leq \wh w_p\leq \frac{\wh g_0}{2p^2} \quad \mbox{for} \quad
  |p|\leq \delta \; .
\label{wp}
\ee

Using the lower bounds \eqref{uniflow} and
the  approximation \eqref{lip}
of $\wh g_p, \wh f_p$ for small $p$, we obtain
the following estimate on $e_{p}$ defined in \eqref{ep}:
\be
     \left\{
\begin{array}{lll}
 |e_p| \le C& \mbox{ for}
&
  \forall p  \\ && \\
e_p \sim  -\frac{\varrho \wh g_0}{2 (p^2+\varrho \wh f_0) }& \mbox{ for}
&
\delta^{-1}\varrho\lambda^2 \leq  |p|\leq \delta \\ &&\\
|e_p| \sim \frac{\varrho |\wh g_p|}{ 2 p^2}&  \mbox{ for } &
|p|\geq \delta \; ,
\end{array}
\right.
\label{apri}
\ee
where $C$ is a constant depending on $V$
and the notation $A\sim B$ indicates that
$A$ and $B$ have the same sign and
$\frac{1}{2} |A| \leq |B| \leq |A|$.
 Here we have used the fact that
$0 < \wh f_0, \wh g_0 < {\widehat V}_0$ and that $\wh f_{0}$
is order ${\widehat V}^{2}$ while $\wh g_{0}={\widehat V}_{0}-\wh f_{0}$.
Similarly,  we have
\be
\label{eq:650b}
   \Big| \frac{e_p}{1-2e_p}\Big|\lesssim
 \left\{
\begin{array}{lll}
\frac{\varrho \wh g_0}{p^2+
  \varrho {\widehat V}_0} & \mbox{ for}
&
|p|\leq \delta \\ &&\\
\frac{\varrho |\wh g_p|}{p^2}&  \mbox{ for } &
|p|\geq \delta
\end{array}
\right.
\ee
and
\be \label{eq:650c}
   \Big| \frac{e_p(1-e_p)}{1-2e_p}\Big|\lesssim
 \left\{
\begin{array}{lll}
\frac{\varrho \wh g_0}{p^2+
  \varrho \wh f_0} & \mbox{ for}
&
|p|\leq \delta \\ &&\\
\frac{\varrho |\wh g_p|}{p^2}&  \mbox{ for } &
|p|\geq \delta\; ,
\end{array}
\right.
\ee
where for positive quantities $A\lesssim B$ indicates that
$A\leq CB$ with a constant $C$ depending only on $V$.

\begin{lemma}\label{lm:E}
Suppose $e_{p}$ is given by \eqref{ep}. Then the energy $E=\langle H \rangle_{\Psi}$
of the state \eqref{state} satisfies
\be
\begin{split}
  E = \; &4\pi aN\varrho \\ &+\sum_{p\ne 0} \Bigg( p^2
  \frac{e_p^2}{1-2e_p} + \varrho {\widehat V}_p  \frac{e_p}{1-2e_p}
 \\ &\hspace{1cm} +\frac{1}{2|\Lambda|} \sum_{r\ne 0} {\widehat V}_{p-r}e_pe_r  + \frac{\varrho^2}{2}
{\widehat V}_p\wh w_p\Bigg) \\ &+ O(N\varrho^2|\log\varrho \, |)
\label{E3}
\end{split}
\ee
as $\varrho\to 0$.
\end{lemma}

\noindent{\bf Proof.} We first prove that $\Omega_{2}+ \Omega_4$ are
 negligible.
Note that, using $|\wh g_p|\leq \wh g_0\leq \wh V_0$,
 and the bounds \eqref{apri}, (\ref{eq:650b}), we have
\begin{align}\label{e2est}
  \frac{1}{|\Lambda|}\sum_{p\neq 0} \frac{e_p^2}{1-2e_p}  \leq \; &\frac{C}{|\Lambda|}
  \sum_{0 \neq
|p|\le \delta} \frac{(\varrho {\widehat V}_0)^2}{p^2(p^2 + \varrho {\widehat V}_0)} \nonumber
\\ &+  \frac{C}{|\Lambda|} \sum_{|p|\ge \delta} \frac{(\varrho \wh g_p)^2}{p^4} \nonumber \\
\leq \; & C\varrho^{3/2}\; .
\end{align}
Similarly, we find
\begin{align}\label{eeest}
    \frac{1}{|\Lambda|}\sum_{p\neq 0}  \frac{|e_p(1-e_p)|}{1-2e_p} \leq \; &
  \frac{C}{|\Lambda|} \sum_{0 \neq |p|\leq \delta}
   \frac{\varrho \wh g_0}{p^2 + \varrho \wh f_0} \nonumber \\ &+ \frac{C}{|\Lambda|} \sum_{|p|\geq \delta}
   \frac{\varrho \wh g_p}{p^2} \nonumber \\ \leq \;& C\varrho 
\end{align}
and
\begin{align}\label{eest}
    \frac{1}{|\Lambda|}\sum_{p\neq 0} \Big| \frac{e_p(1-e_p)}{1-2e_p}\Big|^2 \leq\; &
  \frac{C}{|\Lambda|} \sum_{0\neq |p|\leq \delta}
   \Big(\frac{\varrho \wh g_0}{p^2 + \varrho \wh f_0}\Big)^2
  \nonumber \\ &+ \frac{C}{|\Lambda|} \sum_{|p|\geq \delta}
   \frac{(\varrho \wh g_p)^2}{p^4} \nonumber \\ \leq \; & C\varrho^{3/2}
\end{align}
with constants depending on $V$.
In terms of $c_p$'s, we have, by \eqref{rel},
\be\label{orderc}
\sum_{p\ne 0} \frac{|c_p|^2}{1-|c_p|^2} \le C N \varrho^{1/2}, 
\qquad
\sum_{p\ne 0} \frac{|c_p|}{1-|c_p|^2}\le C N .
\ee 
The lower bounds in \eqref{apri} and \eqref{eq:650b}
also imply that
\be \label{orderclow}
\frac{1}{|\Lambda|}\sum_{p\ne 0} \frac{|c_p|^2}{1-|c_p|^2}
\sim C \lambda^{3/2}\varrho^{3/2}
\big[1+ O(\lambda\varrho)^{1/2}\big]\; .
\ee
In particular, we have  proved \eqref{n1} after recalling \eqref{tot}.

Using \eqref{e2est} and \eqref{eest}, the following terms are negligible from \eqref{E2}:
\begin{align}
 &\frac{1}{2|\Lambda|}\sum_{p\neq 0}\sum_{r\ne 0,\pm p}
({\widehat V}_0+{\widehat V}_{p-r}) \nonumber \\ &\hspace{2cm} \times \frac{e_p^2e_r^2}{(1-2e_p)(1-2e_r)}  \leq  CN\varrho^2
\label{neg0} \\
  & \frac{{\widehat V}_0}{2|\Lambda|} \Big( \sum_{p\ne 0}
 \frac{e_p^2}{1-2e_p}\Big)^2  \leq 
 CN\varrho^2
\label{neg1} \\
&   \sum_{p\ne0}
\frac{{\widehat V}_p+{\widehat V}_0}{|\Lambda|}\Big(\sum_{r\ne 0} \frac{e_r^2}{1-2e_r}\Big)
 \frac{e_p^2}{1-2e_p}  \leq
 CN\varrho^2
\label{neg2} \\
&\sum_{p\ne0}\frac{{\widehat V}_0}{|\Lambda|}
\Big(\frac{e_p(1-e_p)}{1-2e_p}\Big)^2
 \frac{1-2e_p+4e_p^2}{(1-e_p)^2} \leq  C\varrho^{3/2}\; ,
\label{neg3}
\end{align}
where we used
$$
\frac{1-2e_p+4e_p^2}{(1-e_p)^2}  =1+ 3c_p^2 \leq 4
$$
and similarly
\be
\sum_{p\ne0}\frac{{\widehat V}_{2p}}{|\Lambda|}
\Big(\frac{e_p(1-e_p)}{1-2e_p}\Big)^2
 \frac{1-2e_p+2e_p^2}{(1-e_p)^2} \leq C \varrho^{3/2} \,.
\label{neg6}\ee
Notice that the terms \eqref{neg3} and \eqref{neg6} are not extensive.
All constants depend on $V$.
We have thus proved that the $\Omega_{2}+ \Omega_4$ are bounded by
$N \varrho^{2}$.

\bigskip

We now replace ${\widehat V}_p$ by ${\widehat V}_{p-r}$ in the last
term of the main term $E_{M}$ \eqref{E2.5}. The
difference can be estimated by using \eqref{lip} and \eqref{eeest} as
\begin{align}
 \Bigg| &\sum_{p\ne0} \frac{1}{|\Lambda|}
\Big(\sum_{r\ne 0} ({\widehat V}_p-{\widehat V}_{p-r})\frac{e_r^2}{1-2e_r}\Big)
 \frac{e_p(1-e_p)}{1-2e_p}\Bigg| \nonumber\\
& \leq \frac{C}{|\Lambda|}\Bigg( \sum_{0 \neq
|r|\leq \delta}
 \frac{(\varrho {\widehat V}_0)^2|r|}{(r^2+\varrho \wh f_0)(r^2+\varrho {\widehat V}_0)}+
 \sum_{|r|\geq \delta}
 \frac{(\varrho \wh g_r)^2|r|}{r^4}
 \Bigg) \nonumber \\ &\hspace{1cm} \times
\sum_{p\neq 0} \frac{|e_p(1-e_p)|}{1-2e_p} \nonumber\\
& 
\leq CN\varrho^2 |\log\varrho| \; .
\end{align}
We remark that this is the only term of size $N\varrho^2|\log\varrho \, |$ and
is the candidate for the third order term.

After this change, we can combine the two terms in the
last two lines of \eqref{E2.5} as
\begin{align}
 \sum_{p\ne0}\frac{1}{2|\Lambda|}&\sum_{r\ne 0,\pm p} {\widehat V}_{p-r}
\frac{e_p(1-e_p)e_r(1-e_r)}{ (1-2e_p)(1-2e_r)}
\nonumber \\ &- \sum_{p\ne0} \frac{1}{|\Lambda|}
\Big(\sum_{r\ne 0} {\widehat V}_{p-r}\frac{e_r^2}{1-2e_r}\Big)
 \frac{e_p(1-e_p)}{1-2e_p} \nonumber\\
   =&\frac{1}{2|\Lambda|} \sum_{p,r\ne 0} {\widehat V}_{p-r}
  \Big( \frac{e_p(1-e_p)}{1-2e_p} - \frac{e_p^2}{1-2e_p}\Big)
 \nonumber \\ &\hspace{1cm} \times \Big( \frac{e_r(1-e_r)}{1-2e_r} - \frac{e_r^2}{1-2e_r}\Big)
\nonumber\\
& - \frac{1}{2|\Lambda|} \sum_{p,r\ne 0} {\widehat V}_{p-r}
  \frac{e_p^2}{1-2e_p}\frac{e_r^2}{1-2e_r}
 \nonumber \\ &- \frac{1}{2|\Lambda|}\sum_{p\ne 0} ({\widehat V}_{0}+{\widehat V}_{2p})
\Big(\frac{e_p(1-e_p)}{ (1-2e_p)}\Big)^2 \nonumber \;.
\end{align}
The last term comes from compensating for removing
the restriction $p\ne \pm r$.  Similarly to \eqref{neg3},
it is not extensive and thus negligible.
The second term is estimated exactly as \eqref{neg1}
after estimating $|{\widehat V}_{p-r}|\leq {\widehat V}_0$; it
is $O(N\varrho^2)$ and thus  negligible.

Since
$$
\frac{e_p(1-e_p)}{1-2e_p} - \frac{e_p^2}{1-2e_p}=e_p
$$
this concludes the proof of Lemma \ref{lm:E}. $\;\;\Box$

\bigskip

{\bf Proof of Theorem \ref{mainth}.}
Our goal is to minimize the energy given in \eqref{E3} in the parameters $e_p$. This can be
done either directly or
via the following observation which converts the nonlocal term involving
both $e_{p}$ and $e_{r}$ to a local term. For two functions
 $\ph, \psi$ defined on $\Lambda^*$, we denote
\[ (\ph , \wh V * \psi) = \frac{1}{|\Lambda|} \sum_{p \neq 0}
\frac{1}{|\Lambda|} \sum_{r \neq 0} \widehat{V}_{p-r} \, \ph_p \psi_r \; .
\]
Hence
$$
 \frac 1 {|\Lambda|}
 \sum_{p\ne 0}  \frac{1}{|\Lambda|} \sum_{r\ne 0} {\widehat V}_{p-r}e_p e_r
=  (e,{\widehat V}\ast e) \,.
$$
We use the identity
\begin{equation*}\begin{split}
   \frac{1}{2} (e,{\widehat V}\ast e) = \; &\frac{1}{2} \big(e+\varrho \widehat{w},
 {\widehat V}\ast(e+\varrho \widehat{w})\big) -
\varrho (e, {\widehat V}\ast \widehat{w}) \\ &- \frac{\varrho^2}{2}
   (\widehat{w}, {\widehat V}\ast \widehat{w})
\end{split}\end{equation*}
and $\wh g={\widehat V}-{\widehat V}\ast \widehat{w}$ (see \eqref{gdef}).
We can combine the last term
with the last term in \eqref{E3} to obtain
\be
- \frac{|\Lambda| \varrho^2}{2}(\wh w, {\widehat V} \ast \widehat{w})
 + \frac{\varrho^2}{2} \sum_{p\neq 0} {\widehat V}_p\wh w_p =
 \frac  1 4 |\Lambda| \varrho^{2}(\widehat{g}, p^{-2} \widehat{g} )
\ee
where we have used  $\wh g_p =2p^2\wh w_p$. Thus  we
have
\begin{align}
  E= & \; 4\pi aN\varrho \nonumber \\ &+\sum_{p\ne 0} \Bigg( p^2
  \frac{e_p^2}{1-2e_p} + \varrho {\widehat V}_p  \frac{e_p}{1-2e_p}
 \nonumber \\ &\hspace{1cm}  -\varrho \,
({\widehat V}\ast \wh w)_p \, e_p + \frac{\varrho^2\wh g_p^2}{4p^2}\Bigg) \nonumber\\
&+\frac{1}{2|\Lambda|} \sum_{p,r\ne 0}
{\widehat V}_{p-r}(e_p+\varrho \wh w_p)(e_r+\varrho \wh w_r) \nonumber
\\ &+ O(N\varrho^2|\log\varrho\, | )\;.
\label{E4}
\end{align}

{F}rom the definition of $e_{p}$ in \eqref{ep}, we have for $p^{2} \ge \delta^{-1} \varrho$
\be
    e_p
 =    -\frac{\varrho \wh g_p}{ 2(p^2+2\varrho \wh f_p)} +
\varrho^{2} O\Big( \frac {\wh g_{p}^{2}} {p^{4}}\Big)\; .
\label{ep1}
\ee
Therefore, for $p^{2} \ge \delta^{-1}\varrho$, we have
\be\begin{split}
    e_p + \varrho \wh w_{p}
& =  - \frac{\varrho \wh g_p}{ 2(p^2+2\varrho \wh f_p)}+  \frac{\varrho \wh g_p}{ 2p^2}
 + \varrho^{2} O\Big( \frac {\wh g_{p}^{2}} {p^{4}}\Big)
\\ &= \varrho^{2} O\Big( \frac {\wh g_{p}^{2} + |\wh g_{p} \wh f_{p}|} {p^{4}}\Big) \; .
\label{ep2}
\end{split}\ee

We now prove that the term in the last but one line of \eqref{E4}  is negligible.
By \eqref{wp} and \eqref{apri} we have
\begin{align}
 \frac{1}{2|\Lambda|} &\sum_{p,r\ne 0}
 {\widehat V}_{p-r} (e_p+\varrho \wh w_p)(e_r+\varrho \wh w_r) \nonumber \\
&\leq C|\Lambda| \Big(\int_{ p^{2}\le \delta^{-1} \varrho} \frac{\varrho \wh g_0}{p^2}
+ \int_{p^{2}\ge \delta^{-1} \varrho} \varrho^{2}
 \frac {\wh g_{p}^{2} + |\wh g_{p} \wh f_{p}|} {p^{4}}
\Big)^2 \nonumber \\ &\leq
 CN\varrho^2\; . \nonumber
\end{align}

The minimization of the first three terms in the summation over $p \neq 0$ in
 \eqref{E4} is exactly given by Lemma \ref{le-ep}. With the choice \eqref{ep} for $e_{p}$, we have
 proved that the energy satisfies the following estimate:
\be
 E= 4\pi\varrho Na + Q    + O(N\varrho^2|\log\varrho \, |) \; ,
\label{E5}
\ee
where
\be\label{omega}
\begin{split} Q =  \frac{\varrho}{2}\sum_{p\ne0}
 \Bigg[&\sqrt{\Big(\frac{p^2}{\varrho}+2\wh f_p\Big)
\Big(\frac{p^2}{\varrho}+2{\widehat V}_p\Big)} \\
 &-\Big(\frac{p^2}{\varrho}+{\widehat V}_p +\wh f_p\Bigg)
  +\frac{\varrho \wh g_p^2}{2p^2}\Bigg] \; .
\end{split}\ee

We now take the limit $L \to \infty$ and change the summation to integration.
Changing variables, $x=p^2/\varrho$, $\rd^3p=2\pi\varrho^{3/2}x^{1/2}\rd x$
for $x\in (0,\infty)$, then $ Q $ is given by
\be
 Q = \frac{\pi N\varrho^{3/2}}{(2\pi)^3} \int_{0}^{\infty}
F(x, \sqrt { x \varrho}) \rd x \; ,
\label{integr}
\ee
where
$$
F(x, p)=
 \Big(\sqrt{(x+2\wh f_p)(x+2{\widehat V}_p)}-(x+\wh f_p+{\widehat V}_p) +\frac{\wh g_p^2}{2x}\Big)
x^{1/2} \;.
$$
It should be noted that in our formulas, $x$ and $p$ are always  related via the relation
$x=p^2/\varrho$.

With the notation  $\alpha = 2 \wh f_p /x, \;  \beta  = 2 \wh V_{p}/x$ and recalling
$\wh g_{p}={\widehat V}_{p}-\wh f_{p}$ from \eqref{fg} we can write
\begin{equation*}\begin{split}
F(x, p)= \; &
 \Big( \sqrt{ (1+ \alpha)(1+\beta)} \\ &- \left
 [ 1+\frac {\alpha+ \beta} 2  + \frac { (\alpha- \beta)^{2}} 8   \right  ] \Big)
x^{3/2}\; .
\end{split}
\end{equation*}
We divide the integration \eqref{integr} into $x \ge c \varrho^{-1}$ and $x \le c \varrho^{-1}$
regimes
for some small constant $c$.
Since $ |{\widehat V}_p|+|\wh f_p| $ is bounded for
all $p$, in the region $x \ge c \varrho^{-1}$,
we have $|\alpha| + |\beta| \ll 1$ for $c$ independent of
$\varrho$ and $\varrho$ is small enough, a condition we assume from now on.
We can thus  expand $\alpha$ and $\beta$ in Taylor series  and it turns out
that the leading contribution is the third order
term $ (\alpha+\beta)(\alpha-\beta)^{2}/16$.  Hence
we have  $F(x, \sqrt {\varrho x}) \ge 0$ for $x \ge c \varrho^{-1}$
and $\varrho$ small. Furthermore,  we have  the following estimate:
\be\label{8}
 \int_{c \varrho^{-1}}^{\infty}
F(x, \sqrt { x \varrho}) \rd x
\le \int_{c \varrho^{-1}}^{\infty} x^{-3/2}\rd x
\le C \sqrt \varrho \; .
\ee
Similarly, the following inequalities, which will be useful later on,  also hold:
\be\label{8-1}
 \int_{c \varrho^{-1}}^{\infty}
F(x, 0) \rd x
\le C \sqrt \varrho, \qquad F(x, 0) \ge 0  \; \text{for} \; x \ge c \varrho^{-1} \; .
\ee

For $x \le c \varrho^{-1}$, we again use $\wh g_{p}={\widehat V}_{p}-\wh f_{p}$  and
rewrite  $F(x, p) =    x^{-1/2} G(x, p)$
where
\begin{align}\label{8.4}
G(x, p) & =
 \frac{   \wh g_p^2  [ \sqrt{(x+2\wh f_p)(x+2{\widehat V}_p)}+\wh f_p+{\widehat V}_p -x]}{2
\big[\sqrt{(x+2\wh f_p)(x+2{\widehat V}_p)}+(x+\wh f_p+{\widehat V}_p)\big]} \nonumber \\
& =
 \frac{   \wh g_p^2  [ 4(\wh f_p+{\widehat V}_p)x-\wh g_p^2 ]}{2
\big[\sqrt{(x+2\wh f_p)(x+2{\widehat V}_p)}+(x+\wh f_p+{\widehat V}_p)\big]}
\nonumber \\ & \hspace{.5cm} \times  \frac{1}{\sqrt{(x+2\wh f_p)(x+2{\widehat V}_p)}
+x-\wh f_p-{\widehat V}_p}  \; .
\end{align}

The numerator in \eqref{8.4} may vanish only  when
\be
\sqrt{(x+2\wh f_p)(x+2{\widehat V}_p)} = x-\wh f_p-{\widehat V}_p \; .
\ee
Solving for $x$, we have
\be\label{8.2}
x = \frac {(\wh f_p - \wh V_p)^{2}} { 4 (\wh f_p+\wh V_p)}
\ee
In the regime $x \le c \varrho^{-1}$, $ |p| \le \sqrt c$ and from the continuity of $
\wh f_p, \wh V_p$ \eqref{lip}, the leading contribution of $
\wh f_p, \wh V_p$ is given by  $\wh f_{0}, \wh V_{0} > 0$.
Hence for $c$ small enough (depending on $\lambda$, but not
on $\varrho$), the solution  \eqref{8.2} satisfies that
\be\label{8.3}
x = \frac {(\wh f_p - \wh V_p)^{2}} { 4 (\wh f_p+\wh V_p)} < \wh f_p+\wh V_p \; .
\ee
Therefore, the numerator in \eqref{8.4} is positive in the neighborhood of the solution
\eqref{8.2} and is thus  also  positive for all $x \le c \varrho^{-1}$.  As a side remark,
when combined with the previous  argument for  $x \ge c \varrho^{-1}$, this proved that
$G(x, \sqrt{ \varrho x})$ and $F(x, \sqrt{ \varrho x})$ are  positive everywhere.

{F}rom  \eqref{8.4},
$G$ depends smoothly on $x$, $\wh f_p$ and $\wh V_p$ in the regime  $x \le c \varrho^{-1}$.
Using the uniformly Lipschitz continuity of $\wh f_p$ and $\wh V_p$ \eqref{lip},
we thus have
$$
|G(x, p) - G(x, 0)| \le C |p| (1+ x)^{-1} \; .
$$
Here we have used the second line of \eqref{8.4} to obtain the decay in $x$ for $x$ large.
Therefore, we have the error estimate
\begin{equation*}
\begin{split}
 \int_{0}^{c \varrho^{-1}} |F(x, \sqrt {x \varrho})-F(x, 0)| \rd x
 &\le C \sqrt \varrho  \int_{0}^{c \varrho^{-1}} \frac{\rd x}{1+ x}
\\ &\le C \sqrt \varrho \,|\log \, \varrho |\; .
\end{split}
\end{equation*}
Together with \eqref{8} and \eqref{8-1}, the same estimate holds if the integration domain
is extended to the whole $\bR^{+}$.

{F}rom \eqref{hdef},  \eqref{gdef} and \eqref{ag} we see that $h=\wh f_0/\wh g_0$.
 With this notation we have
\begin{equation*}\begin{split}
 Q
 =\; & \frac{\pi N\varrho^{3/2}}{(2\pi)^3} \int_{0}^{\infty} \rd x\, x^{1/2} \\
 &\times \Big(\sqrt{(x+2\wh f_0)(x+2\wh f_0+2\wh g_0)}\\ &
 \hspace{3cm} -(x+2\wh f_0+\wh g_0) +\frac{\wh g_0^2}{2x}\Big)\\
& + O(N\varrho^{2} |\log \varrho |) \\
 = \; & \frac{\pi N\varrho^{3/2}\wh g_0^{5/2}}{(2\pi)^3} \int_{0}^{\infty}
\rd y \, y^{1/2} \\ &\times   \Big(\sqrt{(y+2h)(y+2+2h)}-(y+1+2h) +\frac{1}{2y}\Big)
\\ &+ O(N\varrho^{2} |\log \varrho |) \;.
\end{split}
\end{equation*}
Recall $\wh g_{0}= 8 \pi a$ from \eqref{ag} and the definition of $\Phi(h)$
from \eqref{def:Phi}. We can thus write  $Q$ as
$$
 Q   = 4\pi a N\varrho \Bigg[ \sqrt{\frac{32}{\pi}} \Phi(h) (a^3\varrho)^{1/2}
\Bigg]+ O(N\varrho^{2} |\log \varrho \, |)\; ,
$$
Together with \eqref{E5}, this proves the main Theorem. $\Box$

{\it Acknowledgements.} L. Erd{\H o}s is partially supported by SFB-TR12 of
the German Science Foundation.
B. Schlein is supported by a Sofja-Kovalevskaya Award of
the Humboldt Foundation. He is on leave from Cambridge University, UK.
H.-T. Yau is partially supported by NSF grants DMS-0602038, 0757425 and 0804279.

\end{document}